# Sampling Rate Independent Filtration Approach for Automatic ECG Delineation

H. Chereda, S. Nikolaiev, Y. Tymoshenko
Institute for Applied System Analysis
NTUU "KPI"
Kyiv, Ukraine

*Abstract* — In this paper different types of ECG automatic delineation approaches were overviewed. A combination of these approaches was used to create sampling rate independent filtration algorithm for automatic ECG delineation that is capable of distinguishing different morphologies of T and P waves and QRS complexes.
Created filtration algorithm was compared with algorithme à trous. It was investigated that continuous wavelets transform with proposed automatic adaptation for different sampling rates procedure can be used for delineation problem.

## I. Introduction

The healthcare industry is now on the cusp of disruptive changes and the new technologies are being developed will truly alter the way how the medical care is provided to the patients. The medicine in the 21st century will be functioning in the framework of a fundamentally new P3 paradigm: predictive, preventive and personalized medicine (PPPM). This means that the healthcare will become proactive but not reactive and medical sensors become ubiquitous, the streams of bio-data available to clinicians will completely overwhelm their ability to understand this amount of information and react in real time. To deal with this "ocean of bio-signals" (Big Data), we need to develop fast and reliable automatic signal processing algorithms that can adapt to the peculiarities of individual person. [1]

Nowadays the cardio vascular diseases are the major cause of death and because of that more and more businesses every year create smart-garments that measure electrocardiogram (ECG) in daily life. These garments require ECG automatic real-time analysis and heart pathologies detection software.

So we may state the importance of automatic ECG signal processing and analysis tools development. These tools contain several stages of information processing like raw analog signal noise reduction, digitized signal filtering, digitized ECG delineation for marking of P, T waves and QRS complexes (heartbeats), time- and amplitude- and frequency-based feature retrieval, machine-learning or rule based approaches for pathology detection and prediction. [2] During applying these stages researcher will explore the ECG recordings with completely different sample rates. For example Physionet database contains dozens gigabytes of ECGs recorded with a sampling frequency in the range 250 - 1000 Hz. But current known approaches does not allow to easily compute data with different sampling rates because of tedious and laborious resampling procedure and digital filters responses interpolating. So we propose a novel simplified filters adaptation approach for automatic adjustment to different sampling rates of ECG signals.

The paper is organized as follows: in Section II, we focus on ECG automatic delineation algorithms overview. The usage of continuous wavelet transform (CWT) filtration for the input ECG signal is described in Section III. Section IV is focused on features of stationary wavelet transform (SWT). Procedure for automatic selection of the appropriate scale for CWT is proposed in Section V. The obtained results of our approach and comparison with SWT method are discussed in Section VI. Finally, the conclusions are presented in Section VII.

## II. Delineation Algorithms Overview

All delineation algorithms can be divided into two major groups: those that detect only QRS complexes peaks and wave delineation algorithms that find all peaks including full marking of P and T waves.

One the most famous examples from the first group is Pan-Tompkins approach [3]. It is used to find R-peak position on the raw ECG signal by calculating adaptive thresholds.

We consider the second type of algorithms, where determination of P- and T-waves is performed after defining the location of QRS complexes. We can name Chesnokov [4] and Laguna [5] approaches as great examples of such algorithms. But even their methods are not ideal.

Chesnokov ECG delineator uses CWT that performs digitized ECG signal filtering equally good for given wavelet analyzing frequencies regardless of the ECG sampling rate.

One of the drawbacks of the approach is that the proposed architecture cannot automatically determine whether the T-wave is biphasic. This parameter can be optionally set by the user before starting the delineator, but it requires a prior knowledge from the user about the analyzed ECG. Also this approach cannot differentiate between ascending and descending T-waves.

Laguna delineator approach allows annotating peaks for any configuration of P- and T- waves and QRS-complexes. But the main drawback is that filters used in SWT (algorithme à trous [5]) need to be individually tuned for ECG signals recorded with different sampling rates.

Our goal is to combine these two approaches to create sampling rate independent filtration algorithm for automatic ECG delineation that is capable of recognizing biphasic, ascending and descending T-waves.

### III. METHODS DESCRIPTION

The ECG signal consists of different parts: complexes (heartbeats ranges) and waves (like P- and T-waves). Physicians use them to determine heart pathologies. The waves contain various spectral components present at certain moments of time that can be automatically analyzed. To analyze these components in the time-frequency representation special tools and approaches are used.

One well-known approach is short time Fourier transform (STFT) [6]. The main idea is to apply Fourier transform to parts of the signal (called windows) where the signal appears to be stationary. But the biggest challenge for this approach is to find optimal window width. To overcome this problem the wavelet transform (WT) approach is used in recent years. This approach implements the decomposition of non-stationary signal on the basis obtained by compression and displacement of a function (prototype wavelet). According to [7] a good selection of parent wavelet will allow to obtain a satisfactory resolution for both time and frequency domains.

Formally continuous wavelet transform can be represented as a function of two variables:

$$T_x(a,\tau) = \frac{1}{\sqrt{a}} \int_{-\infty}^{+\infty} x(t)\psi^*(t-\tau)dt, \quad (1)$$

where $a$ is a scale, $\psi(t)$ - prototype wavelet, $x(t)$ is the signal. It can be considered as inner product in $L_2(R)$ (space of square-integrable functions, defined on the real axis) and as a mutual correlation of the signal and the wavelet. The larger scale - the lower frequency is extracted by the CWT, $f \sim 1/a$. Parameter $\tau$ – offset of the wavelet. The asterisk defines a complex conjugation of the wavelet.

Equation (1) could give the impression that an exact value of the signal frequency can be extracted at a certain moment of time. However, taking into account a broad interpretation of Heisenberg uncertainty principle, in general case this conclusion is not true. From this principle the fact follows - it is impossible to determine which harmonic signal components are present in a fixed time, you can only get an idea of a certain frequency range at a certain time interval.

If prototype wavelet $\psi(t)$ is the derivative of some smoothing function $\theta(t)$ then CWT of signal $x(t)$ at scale $a$ is [8]:

$$T_x(a,\tau) = -a\left(\frac{d}{d\tau}\right) \int_{-\infty}^{+\infty} x(t)\theta_a(t-\tau)dt, \quad (2)$$

where $\theta_a(t) = (1/\sqrt{a})\theta(t/a)$ is the scaled version of the smoothing function. The CWT at fixed scale $a$ is proportional to the derivative of the filtered signal with a smoothing function $\theta_a(t)$. Zero-crossings of the WT correspond to the local maxima or minima of the smoothed signal at different scales, and the maximum absolute values of the wavelet transform are associated with maximum slopes in the filtered signal. A quadratic spline wavelet which is a derivative of a smoothing function is used in this work.

### IV. DEPENDENCY BETWEEN FREQUENCY, SAMPLING RATE AND SCALE

Fixed scale CWT extracts certain interval of frequency components at a fixed sampling rate. The analyzed by CWT frequency $f$ at a fixed scale $a$, is proportional to the sampling rate $s_r$. These values are bounded by the following equation:

$$f = f_c \cdot s_r / a \quad (3)$$

To obtain the equality, we need to find a non-dimensional constant $f_c$. For a certain wavelet at fixed scale $a$ numerical value of the constant equals to the frequency, where Fourier spectrum of the wavelet reaches it's maximum. We established that $f_c = 0.2685$ for quadratic spline wavelet:

$$\psi(x) = \begin{cases} 0, x < 0 \wedge x \geq 4 \\ 2x^2, x \geq 0 \wedge x < 1 \\ -6x^2 + 16x - 8, x \geq 1 \wedge x < 2 \\ 6x^2 - 32x + 40, x \geq 2 \wedge x < 3 \\ -2x^2 + 16x - 32, x \geq 3 \wedge x < 4 \end{cases} \quad (4)$$

and $f_c = 0.16$ for derivative of a Gaussian smooth function.

$$\psi(x) = -xe^{-x^2/2}. \quad (5)$$

The scale $a$ and the offset $\tau$ of the wavelet can be discretized. This allows us to use the main idea of discrete wavelet transform (DWT) - decomposition of the $y(t) \in L_2(R)$ in approximating and detailing parts ($2^j \in Z$ represents the scale factor):

$$y_{j+1}(t) = \sum_{k \in Z} a_{j,k}\varphi_{j,k}(t) + \sum_{k \in Z} d_{j,k}\psi_{j,k}(t), \quad (6)$$

where $\varphi_{j,k}(t) = \sqrt{2^j}\varphi(2^j t - k)$ - is the scale function; $\psi_{j,k}(t) = \sqrt{2^j}\psi(2^j t - k)$ – wavelet function.

The DWT can be implemented by passing the discrete time signal through a high pass and a low pass filters. The original signal can be obtained by reconstruction filter bank.

The authors in [5] have used the analysis filter bank based on quadratic spline wavelet (4) in SWT:

$$h[n] = 1/8 \cdot \{\delta[n+2] + 3\delta[n+1] + 3\delta[n] + \delta[n-1]\}$$
$$g[n] = 2 \cdot \{\delta[n+1] - \delta[n]\}. \quad (7)$$

The equivalent frequency response for the filters in SWT for k-th scale is

$$Q_k(e^{j\omega}) = \begin{cases} G(e^{j\omega}), k = 1 \\ G(e^{j2^{k-1}\omega}) \cdot \prod_{l=0}^{k-2} H(e^{j2^l\omega}), k \geq 2 \end{cases} \quad (8)$$

where $H(e^{jw})$, $G(e^{jw})$ are the frequency responses of filters (8).

$$H(e^{j\omega}) = e^{j\omega/2}(\cos\frac{\omega}{2})^3$$
$$G(e^{j\omega}) = 4je^{j\omega/2}(\sin\frac{\omega}{2}). \quad (9)$$

This filter bank was used with the sampling frequency equal to 250 Hz. For the adaptation to other sampling rates, the authors in [5] adequately resample the equivalent filter impulse responses at 250 Hz to other sampling rate values. But this procedure is time-consuming and can't be done automatically for all sampling rates.

## V. Automatic Selection of the Appropriate Scale for CWT Depending on the Sampling Rate

The SWT allows computing the signal decomposition into certain frequency bands, but only at fixed sampling rate.

So, we propose to define frequencies, analyzed by SWT at scales $2^1$, $2^2$, $2^3$, $2^4$, $2^5$ concerned to 250 Hz sampling rate. The same frequencies will be analyzed in a signal by our filtration approach for any sampling rate $s_r \geq 250$ Hz. Scales would be recalculated depending on the sampling rate for these purposes. The procedure of finding an appropriate scale for CWT is described below.

On the first step the determination of the frequencies which are analyzed by SWT responses (8) at scales $2^1$ - $2^5$ is performed. Each of these frequencies corresponds to the point, where the maximum of the SWT frequency response is achieved (Fig. 1).

On the second – the obtained frequency values can be found in the Table I. These frequencies are used as constants for wavelet scale recalculation depending on the sampling frequency.

After that on the third step the scale $a$ can be defined as a function of sampling rate $s_r$:

$$a(s_r) = s_r \cdot f_c / f. \quad (10)$$

By this way we obtain an adaptation for the scale for any sampling rate at fixed frequency $f$ taken from the Table I analyzed by the wavelet.

TABLE I. ANALYZED FREQUENCIES FOR GIVEN SCALES

| Scale, k | Frequency $f$, Hz |
|---|---|
| $2^1$ | 125 |
| $2^2$ | 36.90 |
| $2^3$ | 17.17 |
| $2^4$ | 8.43 |
| $2^5$ | 4.21 |

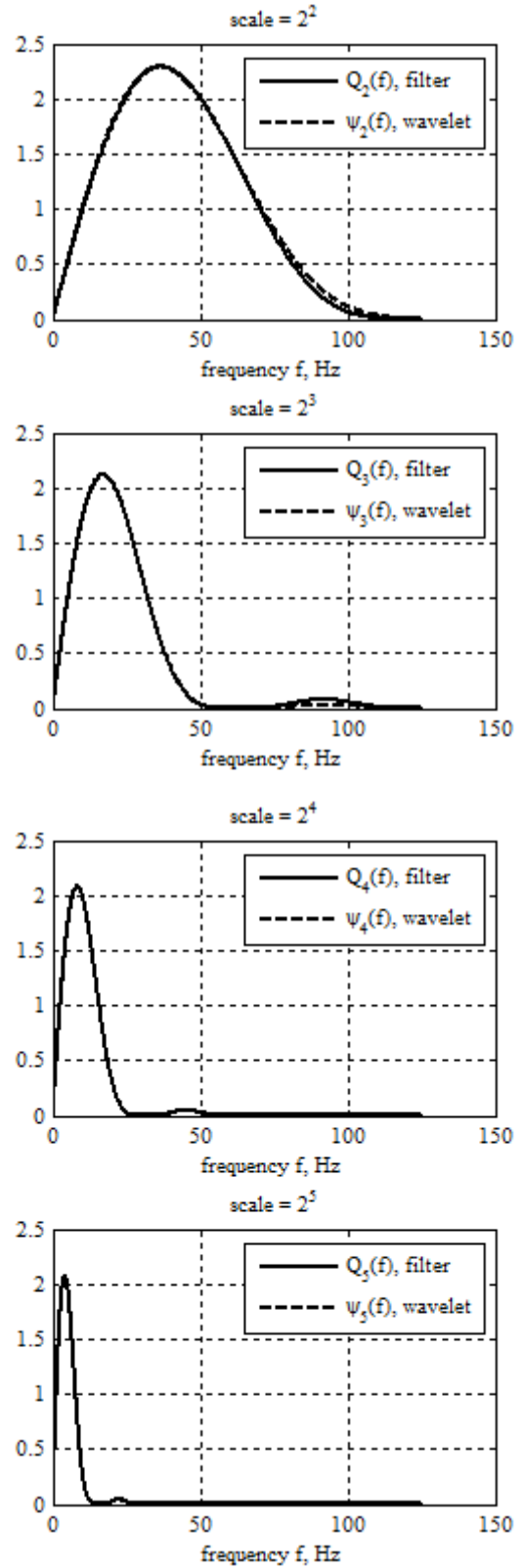

Fig. 1. Comparison of frequency responses for SWT and CWT based filtration at different scales.

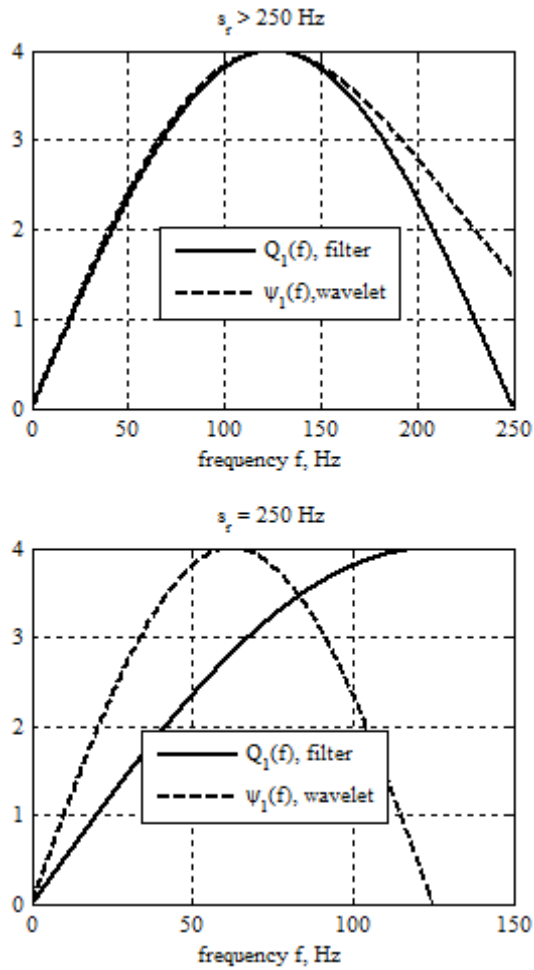

Fig. 2. Comparison of frequency responses for SWT filters and CWT based filtration at $2^1$ scale.

Let's compare the CWT frequency response and the frequencies responses from (8). For this purpose we compute Fourier spectrum of the wavelet (4) assuming that the wavelet's scale was recalculated in accordance with (10). The frequency responses of the wavelet (4) and filters (8) at k-th scale are denoted respectively as $Q_k(f)$ and $\psi_k(f)$ (Fig. 1).

As for the scale $2^1$, we also have a good approximation at $s_r > 250$ Hz (Fig. 2).

TABLE II. CHEBYSHEV ERRORS AT DIFFERENT SCALES

| Scale, k | Error |
|---|---|
| $2^1$ | 0.50 |
| $2^2$ | 0.35 |
| $2^3$ | 0.27 |
| $2^4$ | 0.16 |
| $2^5$ | 0.086 |

It can be observed that the frequency responses of the quadratic spline wavelet with scales recalculated by (10) constitute precise approximation of the original filters up to a frequency of 125 Hz (or 60 Hz when $s_r = 250$ Hz).

VI. RESULTS

The SWT and CWT methods were applied to ECG signals with 250 Hz sampling rate. Tests were performed on sell100.dat file from QTDB (Fig. 4). We denote signals obtained at a fixed scale $2^k$ by the SWT and CWT (for CWT the scale and corresponding frequency are taken from Table I) as $Q_k[n]$ and $T_k[n]$ respectively. The CWT and SWT methods were compared using of Chebyshev error:

$$e_k = \max_{i=1,N} |Q_k[i] - T_k[i]| \quad (11)$$

But before calculating errors in (11), the signals filtered by CWT and SWT must be normalized in the range of [-1, 1]. For signal $x[n]$:

$$\bar{x}[i] = 2 \cdot \frac{x[i] - \min(x)}{(\max(x) - \min(x))} - 1, i = 1..N \quad (12)$$

The results of comparison can be observed in Table II.

According to Table II, CWT significantly deviates from SWT at scale $2^1$. As we can see from the Fig. 3, the most significant errors are present at spikes of the CWT, but in general, approximation could be interpreted as satisfactory for 125 Hz and 250 Hz sampling rates.

For the third scale $2^3$ and the fifth $2^5$ the obtained approximation is better (Fig. 4, Fig. 5).

Fig. 4 and 5 show that the errors on spikes are decreasing at bigger scales. At the fifth $2^5$ scale we have almost perfect approximation.

Scales $2^1$, $2^2$, $2^3$, $2^4$ are relevant for calculating the thresholds that define the existence of QRS complex.

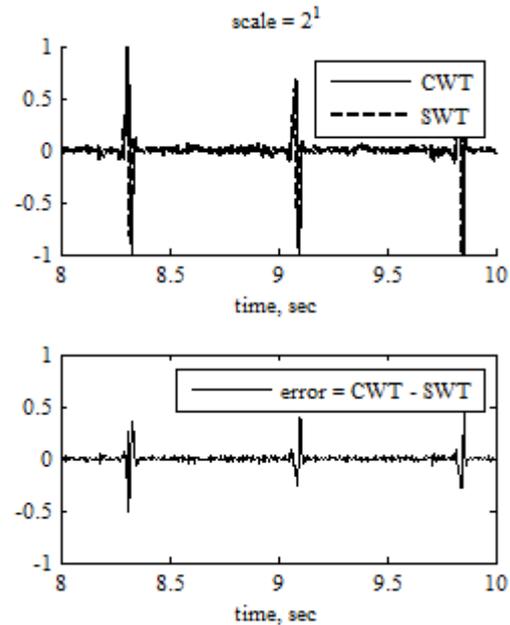

Fig. 3. Wavelet transforms and their error at scale $2^1$.

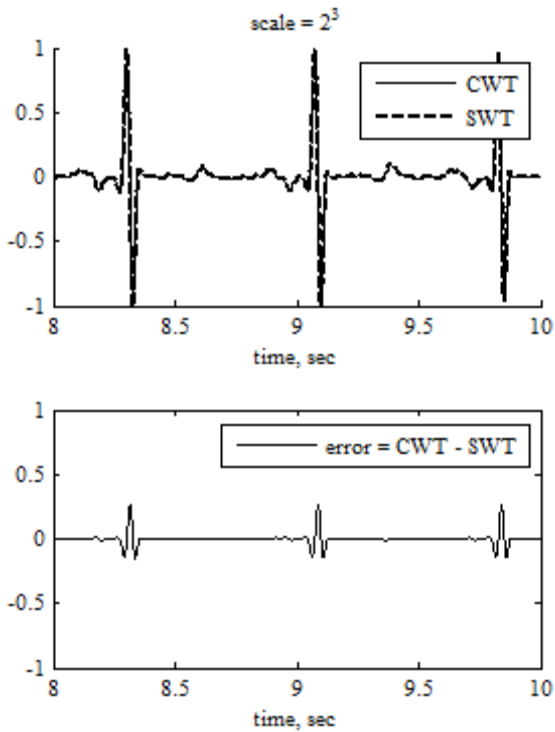

Fig. 4. Wavelet transforms and their error at scale $2^3$.

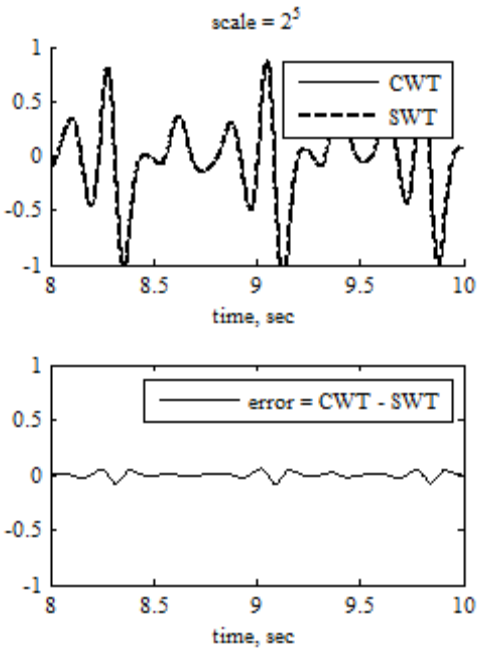

Fig. 5. Wavelet transforms and their error at scale $2^5$.

Despite of obtained deviations between CWT and SWT at the QRS complexes locations at scales $2^1$ and $2^2$ these errors are irrelevant for finding positive maximum and negative minimum pairs (so called maximum modulus lines) on the CWT and SWT graphs (Figs. 3, 4, 5) because maximum modulus lines and (2) guarantees zero crossing of WT with OX axis at local maxima or minima on the ECG signal.

Scales $2^4$ and $2^5$ are used for P, T waves delineation. As for these scales, we obtained a good approximation. More significant Chebyshev error appears at QRS complex location, which is not involved in P and T wave delineation. In this case, there is no need to change the delineation thresholds in Laguna algorithm.

VII. CONCLUSION

In this paper the approach for signal filtering in the ECG automatic delineation problem was researched.

A combination of Chesnokov and Laguna approaches was used to create sampling rate independent filtration algorithm for automatic ECG delineation that is capable of distinguishing different morphologies of T and P waves and QRS complexes.

The filtration algorithm accuracy was investigated for different wavelet scales and despite of big Chebyshev error between CWT and SWT it had no significant influence on the ECG delineation precision. Delineation results for T and P waves are identical to results from [5] in case of tedious and laborious filter responses interpolation procedures are performed.

It was shown that continuous wavelets transform with automatic adaptation for different sampling rates can be used instead of SWT for delineation problem. More over the use of CWT does not require time consuming resampling operations for filter responses as in SWT.